\documentclass[%
reprint,
amsmath,amssymb,
aps,
]{revtex4-2}

\usepackage{graphicx}
\usepackage{dcolumn}
\usepackage{bm}
\usepackage{bbm} 
\usepackage{xcolor}
\usepackage{amsmath}
\usepackage{enumerate}
\usepackage{hyperref}
\usepackage{float}

\newcommand{\Bra}[1]{\left\langle #1 \right|}
\newcommand{\Ket}[1]{\left| #1\right\rangle}

\newtheorem{theorem}{Theorem}

\begin{document}
	
	\preprint{APS/123-QED}
	
	\title{Photonic resource state generation from a minimal number of quantum emitters}
	
	\author{Bikun Li} 
	\email{libk@vt.edu}
	\affiliation{Department of Physics, Virginia Tech, Blacksburg, Virginia 24061, USA}
	
	\author{Sophia E. Economou}
	\email{economou@vt.edu}
	\affiliation{Department of Physics, Virginia Tech, Blacksburg, Virginia 24061, USA}
	
	\author{Edwin Barnes}
	\email{efbarnes@vt.edu}
	\affiliation{Department of Physics, Virginia Tech, Blacksburg, Virginia 24061, USA}

	\date{\today}
	
	\begin{abstract}
		Multi-photon entangled graph states are a fundamental resource in quantum communication networks, distributed quantum computing, and sensing. These states can in principle be created deterministically from quantum emitters such as optically active quantum dots or defects, atomic systems, or superconducting qubits. However, finding efficient schemes to produce such states has been a long-standing challenge. Here, we present an algorithm that, given a desired multi-photon graph state, determines the minimum number of quantum emitters and precise operation sequences that can produce it. The algorithm itself and the resulting operation sequence both scale polynomially in the size of the photonic graph state, allowing one to obtain efficient schemes to generate graph states containing hundreds or thousands of photons.
	\end{abstract}
	
	\maketitle
	
	
\section{Introduction}\label{sec:intro}

Entanglement is widely recognized as playing a critical role in quantum computation, error correction, communication, and sensing. A family of entangled states that features prominently in these applications are graph (or cluster) states. They are key resources in one-way quantum computing paradigms \cite{Raussendorf1WQC,bartolucci2021fusionbased} and in quantum error correction \cite{Schlingemann2001,ShorCode1995, KITAEV20032, nielsen_chuang_2010}. In addition, many quantum repeater schemes \cite{Briegel1998, Dur1999, Sangouard2011, Azuma2015, Muralidharan2016} and quantum sensing protocols \cite{Gottesman2012,Degen2017} rely on graph states. Photonic graph states are especially important because photons are the predominant platform for measurement- and fusion-based computing, and, as flying qubits, they are the only viable choice for quantum networks \cite{Gisin2007} and quantum imaging \cite{Lugiato_2002,Dowling2008}.

Unfortunately, creating photonic resource states is fundamentally difficult. Because photons do not interact with each other, most attempts have focused on probabilistic generation schemes using linear optics and postselection \cite{Browne2005}, which are very resource-intensive, severely limiting the size of the resulting states \cite{Gao2010,Li2020}. This bottleneck can in principle be overcome by instead using a deterministic approach in which entangled photons are produced directly from quantum emitters (i.e., matter qubits). One possibility would be to prepare a graph state on emitters~\cite{Nemoto_PRX_2014,Choi_npjQI_2019} and transduce it to photons, but this requires a number of emitters equal to the size of the target photonic graph state. This daunting resource overhead can be avoided by instead using sequential generation schemes. Refs.~\cite{Schon2005PRL,Schon2007PRA} put forward such an approach that works well for one-dimensional (1D) graph states~\cite{Lindner_2009PRL} and has led to experimental demonstrations~\cite{Schwartz2016,Besse_NatCommun_2020}. However, in the general case where the entanglement structure is more complicated, this method scales exponentially in the size of the target state and can lead to long generation circuits, motivating the search for more efficient approaches.  Refs.~\cite{Economou_2010PRL,Gimeno-Segovia2019} put forward protocols for 2D lattice graphs that leverage the principle that entangled emitters can emit entangled photons. This idea was extended further to develop protocols that deterministically generate resource states for quantum repeaters \cite{Buterakos_2017PRX,Russo2018,Hilaire2021,Zhan2020}---tailored to color centers in Refs.~\cite{Borregaard2020,michaels2021multidimensional}---and one-way computing \cite{Pichler2017,Russo_2019}. Refs.~\cite{Pichler2017,Zhan2020} allowed for the re-interference of photons with emitters to further enhance flexibility in entanglement creation.

Despite this progress and the intense interest this approach has generated among experimentalists, existing graph state  generation protocols are limited to a small subset of graphs or require a number of emitters that scales linearly with the graph size \cite{NJP_9_204_2007,Russo_2019}. This is extremely resource-intensive, especially in light of the schemes for generating repeater graph states presented in Refs.~\cite{Buterakos_2017PRX,Hilaire2021}, which require only two emitters regardless of the number of photons. The required resources (number of emitters and entangling gates) is a critical factor that determines the practical feasibility of the protocol. For a general graph state, finding resource-efficient generation protocols in polynomial time remains an open problem.

Here, we address this challenge by presenting a general approach to generating arbitrary photonic graph states from quantum emitters. Given a target graph state, we show how to determine in polynomial time both the minimal number of emitters required to create it and an explicit generation protocol. The latter consists of a sequence of gate operations and measurements performed on the emitters. Moreover, our protocol naturally takes into account the order in which photons should be emitted, which can be an important consideration for applications, as it is generally preferable to emit photons in the order they are measured to avoid photon storage. Our method provides a recipe for doing this. The broad applicability of our method, its practical relevance, and its efficient use of resources make it ideally suited to the generation of any photonic graph state from various types of quantum emitters.

\begin{figure*}[t]
	\centering
	\includegraphics[width=0.8\linewidth]{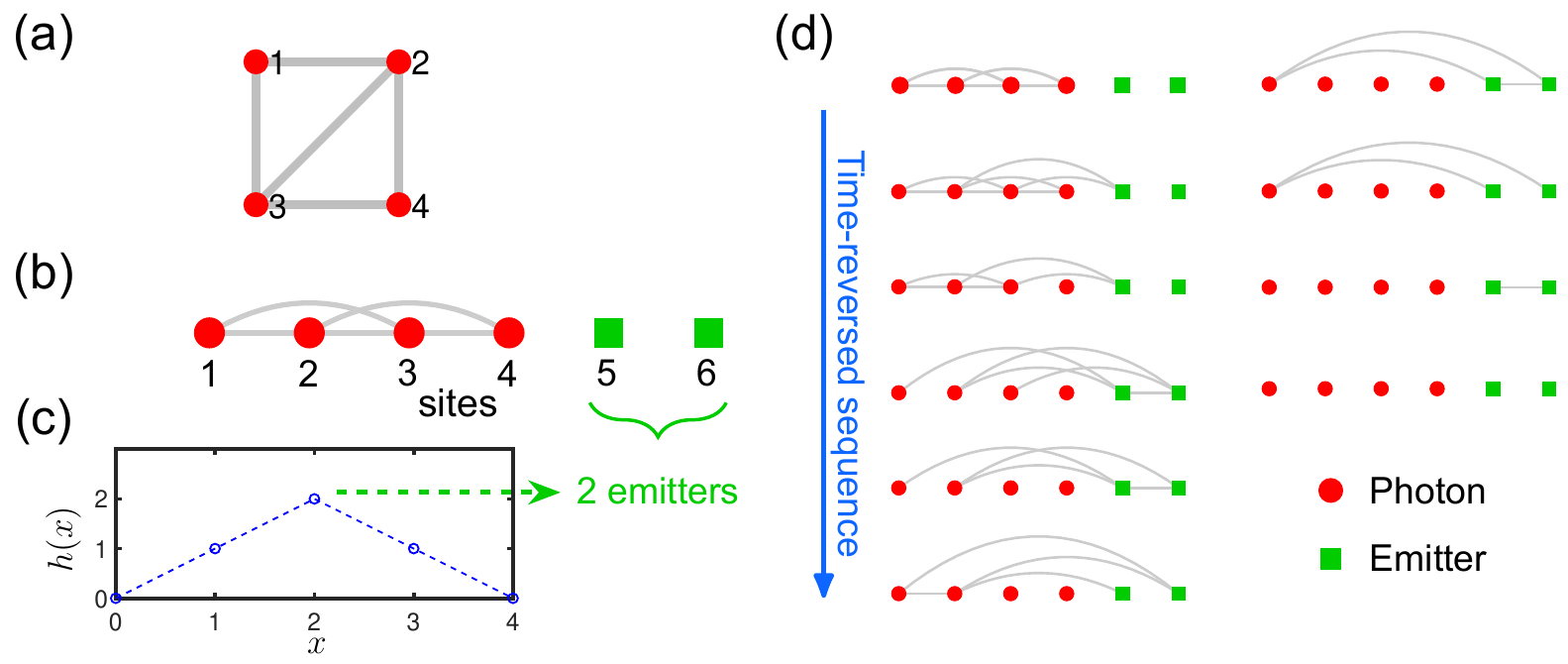}
	\caption{\textbf{Illustration of the protocol solver algorithm.} (a) An example of a 4-photon graph state. (b) The graph is mapped to a 1D lattice. (c) The height function is computed and found to have maximum value 2, implying 2 emitters are needed. These are added to the 1D lattice. (d) Starting from the target state and decoupled emitters, a time-reversed sequence of emitter gates, photon absorption events, and time-reversed emitter measurements is constructed, until all qubits are disentangled. Further details about this example can be found in the Supplementary Information.}
	\label{fig:fig1new}
\end{figure*}

\section{Results and Discussion}\label{sec:results}

\subsection{Overview of the algorithm}\label{sec:overview}

Determining how to efficiently generate an arbitrary photonic graph state from a set of quantum emitters is highly nontrivial and markedly distinct from the problem of finding an efficient quantum circuit that creates a target state on a register of qubits \cite{HMP2006}. Several additional challenges arise in the former, including the fact that qubits are both created and removed, and that different types of qubits (photons vs. emitters), with different roles and allowed gates, are involved. Depending on the experimental setup, there may also be further restrictions, e.g., emitted photons cannot interact with any other qubits following their emission (although schemes that re-interfere photons with emitters have been proposed \cite{Pichler2017,Zhan2020}). Our method addresses these challenges by leveraging three main ingredients: the notion of the height function (which is related to the entanglement entropy), the stabilizer formalism, and the concept of time-reversed emission events and measurements, which we introduce here. 

The first insight is to utilize the so-called height function, which is the entanglement entropy of the system as a function of the partition point when the system is arranged in a 1D lattice and partitioned into two subsystems 
\cite{PhysRevX.7.031016_Nahum,PhysRevB.100.134306}. This function provides information about the entanglement structure of the target state as well as the number of emitters required to produce it. The latter is equal to the maximum value of the height function (see below), which depends on the photon emission order. Optimizing this order is NP-hard in general, although we show that heuristic approaches exist for more structured graphs. Moreover, the height function plays a crucial role in determining the sequence of operations (gates and measurements) needed to generate the target graph state from the emitters.

A second key ingredient is the use of gates from the Clifford group. Given that arbitrary graph states can be generated solely with Clifford gates \cite{VandenNest2004_PRA,Hein2006}, which were also exclusively used in the protocols of Refs. \cite{Lindner_2009PRL,Schwartz2016,Economou_2010PRL,Buterakos_2017PRX,Russo2018,Russo_2019,Gimeno-Segovia2019,Hilaire2021}, restricting ourselves to this set does not affect the generality of our approach. Clifford gates enable the use of the stabilizer formalism, such that we can manipulate Pauli operators instead of keeping track of the whole state. This makes the problem of finding the emission operation sequence tractable, reducing it from exponential to polynomial scaling due to the Gottesman-Knill theorem \cite{Gottesman:1998hu}.

A final key element in our algorithm is that we time-reverse the emission sequence. That is, we start from a target multi-photon graph state and an appropriate number of decoupled emitters (obtained from the height function for the target state), and we determine a sequence of emitter gates, ``time-reversed measurements", and ``photon absorption" events such that the target state is converted to a product state. This is somewhat reminiscent of disentangling circuits used for quantum state tomography of 1D systems~\cite{Cramer_NatCommun_2010}. The final state is a product state because, without loss of generality, photons that have not yet been emitted can be described by qubits prepared in the computational basis state $\Ket{0}$. Photon emission is then modeled as a two-qubit photon-emitter gate that brings the photon from $\Ket{0}$ into an entangled state with the emitters \cite{Lindner_2009PRL}. Because the photon absorption steps are time-reversed versions of photon emission, these too are described by photon-emitter gates.

The run time of the protocol solver algorithm scales as ${\cal O}(n_p^4)$, where $n_p$ is the number of photons in the target graph state. This is a direct consequence of the fact that the algorithm is based on the stabilizer formalism (see Methods section). This is in contrast to previous methods \cite{Schon2005PRL,Schon2007PRA}, which scale exponentially in $n_p$ due to the need to perform singular value decompositions repeatedly. We also show that the number of gates in the final emission sequence scales at most as ${\cal O}(n_p^2)$ (see Methods). However, this assumes two-qubit gates can be applied between any pair of emitters. If this is not the case, then additional SWAP operations are needed, bringing the gate count up to ${\cal O}(n_p^3)$. Therefore, both the protocol solver and the resulting gate sequence it obtains scale polynomially in the size of the target graph state.

Now we provide a more detailed description of the protocol solver algorithm. We begin with a target graph state $\Ket{\psi_p}$ of $n_p$ photons and $n_e$ decoupled emitters, so that the total state is $\Ket{\Psi}=\Ket{\psi_p}\otimes\Ket{0}^{\otimes n_e}$. An $n_p=4$ photon example graph is shown in Fig.~\ref{fig:fig1new}(a). This is what the state of the total system should be at the end of the generation sequence. $n_p$ is set by the size of the desired photonic graph state $\Ket{\psi_p}$, while $n_e$ remains to be determined. We assume the graph representing $\Ket{\psi_p}$ is connected; if this is not the case, then the algorithm can be run separately for each connected subgraph. The state $\Ket{\Psi}$ is fully described by a set of $n=n_p+n_e$ stabilizers $g_m$, $m=1,\ldots,n$, defined such that $g_m\Ket{\Psi}=\Ket{\Psi}$. The full set of $n$ qubits can be arranged in a 1D lattice with site index $x\in\{0,1,2,\ldots,n\}$ (see Fig.~\ref{fig:fig1new}(b)). Sites $x=1,\ldots,n_p$ correspond to the photons and are ordered according to the desired photon emission ordering, while the sites $x=n_p+1,\ldots,n$ are the emitters. The additional $x=0$ site is included as a matter of convention. We can now define the height function $h(x)=S_A$ to be the bipartite entanglement entropy when the 1D lattice is divided into the subregion $A=\{1,2,\ldots,x\}$ and its complement. Note that $S_A=\frac{1}{1-\alpha}\log_2\mathrm{Tr}(\rho^\alpha_A)$ can be any of the R\'enyi entropies; for stabilizer states, they are all equal \cite{Hein2004PRA}. In Ref.~\cite{Schon2005PRL}, it was shown that the state of the emitted photons, $\Ket{\psi_p}$, can be represented by a matrix product state (MPS) with bond dimension $2^{n_e}$. Because the entanglement entropy of a MPS is given by the base-2 logarithm of the bond dimension \cite{Orus2014}, it follows that $n_e$ is equal to the maximum value of $h(x)$. The height function for the graph in Fig.~\ref{fig:fig1new}(a) is shown in Fig.~\ref{fig:fig1new}(c). In this example, its maximum is 2, implying 2 emitters are needed. In general, the maximum of the height function is in fact the minimal number of emitters capable of generating the target graph state, as fewer emitters would be insufficient to match the bond dimension of any exact MPS representation.

The height function can be computed efficiently from the stabilizers. Because products of stabilizers are also stabilizers, there are many equivalent choices for the set $\{g_m\}$. Here, we focus on a particular choice of the stabilizers that we refer to as the echelon gauge \cite{Audenaert_2005NJP}, in which the stabilizer matrix has a row-reduced echelon form (see Methods). When the $g_m$ are in this gauge, the height function can be expressed as \cite{Audenaert_2005NJP}
\begin{equation}\label{eq:height_function}
	h(x)=n - x - \#\{g_m|\verb|l|(g_m)>x\},
\end{equation}
where $\verb|l|(g_m)$ is the index of the left-most (smallest index) site on which $g_m$ acts nontrivially. The last term in Eq.~\eqref{eq:height_function} counts the number of stabilizers that act nontrivially only on sites to the right of (i.e., larger than) $x$. Although Eq.~\eqref{eq:height_function} depends on $n_e$, this dependence cancels out for states like $\Ket{\Psi}$ in which the emitters are decoupled. Therefore we can obtain $n_e$ from the maximum of $h(x)$ on the photonic sites, using only the stabilizers of $\Ket{\psi_p}$.

Once we have the number of emitters $n_e$, we can run the protocol solver algorithm to determine the sequence of gates, time-reversed measurements, and photon absorption events needed to transform the target state $\Ket{\Psi}$ into the initial state $\Ket{0}^{\otimes n}$, which corresponds to decoupled emitters and no photons. We first introduce a photon index $j$ and initialize it to $j=n_p$. The algorithm then consists of four steps:
\begin{enumerate}[(i)]
	\item Transform the stabilizers $g_m$ into echelon gauge if they are not already, then compute the height function $h(x)$.
	\item If $h(j) \ge h(j-1)$, skip to step (iii). Otherwise apply a time-reversed measurement and update the $g_m$ accordingly. 
	\item Apply a photon absorption operation on the $j$-th photon and update the $g_m$ accordingly. If $j>1$, then set $j\to j-1$ and go to step (i). Otherwise, go to step (iv).
	\item All photons are now in state $\Ket{0}$. Apply a series of gates on the emitters to disentangle them, bringing the total state to $\Ket{0}^{\otimes n}$.
\end{enumerate}
This algorithm involves repeated applications of two basic operational primitives: time-reversed measurement and photon absorption. 
During the algorithm, the height function of the current state tells us which of these we need to perform next to bring the state closer to $\Ket{0}^{\otimes n}$. Each photon absorption step disentangles one photon qubit from the rest, starting with the last-emitted photon, $j=n_p$, and working down to the first photon, $j=1$. For our 4-photon example, the graphs at intermediate steps of the algorithm are shown in Fig.~\ref{fig:fig1new}(d). A step by step explanation of this example is given in the Supplementary Information. When the algorithm concludes, we can reverse the entire sequence to obtain an operation sequence that generates $\Ket{\psi_p}$ starting from $n_e$ decoupled emitters. We now describe each of the two operational primitives in more detail, the precise gates they introduce into the generation sequence, and their connection to the height function.

\begin{figure*}[t]
	\centering
	\includegraphics[width=1.0\linewidth]{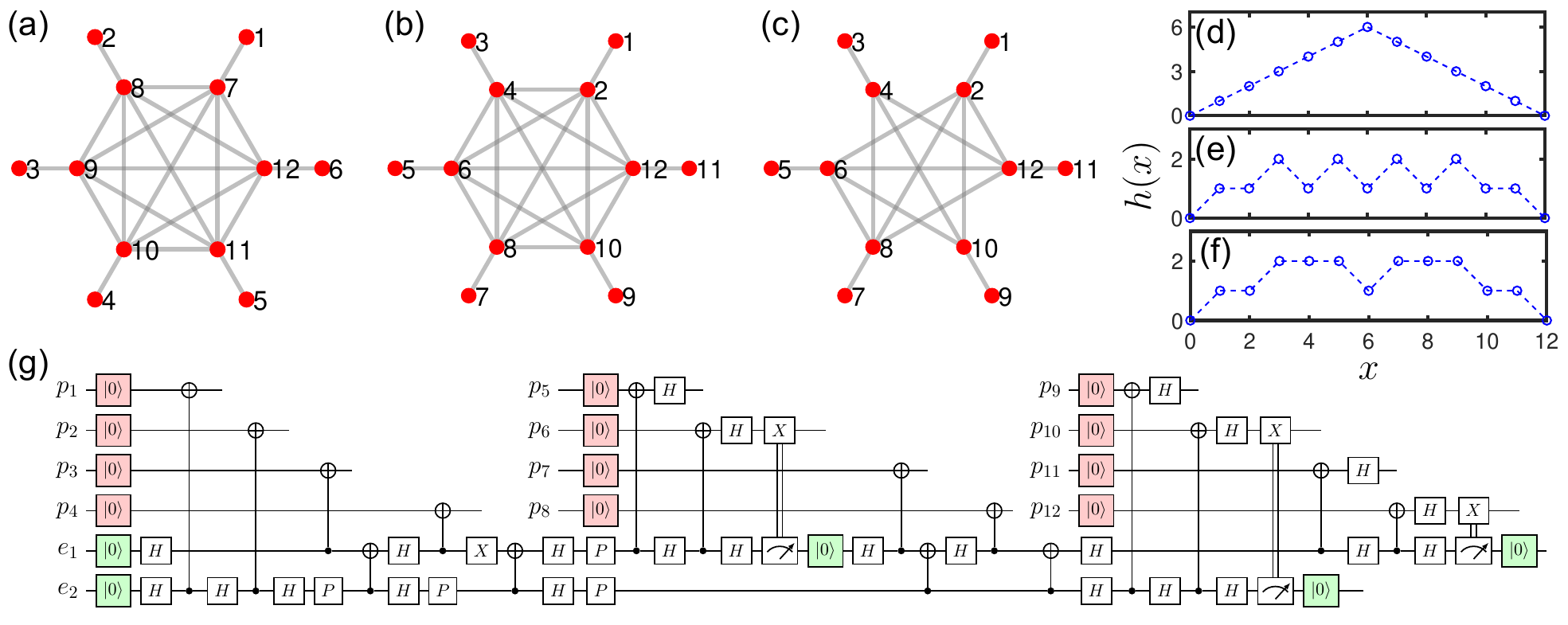}
	\caption{\textbf{Results for repeater graph states.} (a) 12-photon repeater graph state in which external photons are emitted first. (b) Same graph state as in (a), but with ``natural" emission ordering. (c) Same graph state as in (b) but with some unnecessary edges deleted. (d), (e) and (f) show the height functions of the states in (a), (b) and (c), respectively. (g) Emission circuit for state shown in (c), where $H$ is the Hadamard gate, $P=\mathrm{diag}(1,i)$ is the phase gate, and $X\equiv \sigma^x$.}
	\label{fig:RGSs}
\end{figure*}

Photon absorption of the $j$-th photon refers to a time-reversed version of photon emission. For concreteness, we focus on the case where emission is described by a CNOT gate between the photon and its emitter (with the emitter as the control), as in Ref.~\cite{Lindner_2009PRL}, although our algorithm can be adapted to any Clifford gate describing photon emission. Mathematically, the task of absorbing photon $j$ requires finding a stabilizer $g_a$ that can be transformed to $\sigma_j^z$ by applying CNOT$_{ij}$, where $i$ is an emitter site.  
It is possible to find such a stabilizer when $h(j)\ge h(j-1)$. From Eq.~\eqref{eq:height_function}, we see that this condition implies there must be at least one stabilizer, $g_a$, such that $\verb|l|(g_a)=j$. This stabilizer has the form
\begin{equation}\label{eq:ga}
	g_a = \sigma^\alpha_j \sigma_{i_1}^{\beta_1}\cdots \sigma_{i_s}^{\beta_s},
\end{equation}
where $\alpha,\beta_k\in\{x,y,z\}$ label the nontrivial Pauli operators, and $1\le j \le n_p <i_1<\cdots<i_s\le n$. Note that we can assume $g_a$ acts trivially on all photons with index larger than $j$ since these have already been decoupled at this point in the algorithm. We also assume that $g_a$ acts nontrivially on at least one emitter site; if this is not the case, then photon absorption is unnecessary since the $j$-th photon is then already disconnected. To transform $g_a$ into $\sigma_j^z$, we can first apply a local Clifford operation on the $j$-th site and general Clifford operations on the emitters to transform $g_a\to\sigma_j^z\sigma_i^z$, where $i>n_p$ is an emitter site. This can be done for example by applying local Clifford operations to transform $g_a$ to $\sigma^z_j \sigma_{i_1}^{z}\cdots \sigma_{i_s}^{z}$, and then applying CNOT gates on pairs of emitters to transform this to $\sigma_j^z\sigma_i^z$. Applying CNOT$_{ij}$ brings this to $\sigma_j^z$, completing the absorption of the $j$-th photon. Note that we can choose any emitter to absorb the photon; typically, the emitter that requires the shortest circuit to transform $g_a$ into $\sigma_j^z$ is preferred. The resulting circuit is included in the time-reversed generation sequence.

Time-reversed measurements are applied whenever $h(j)<h(j-1)$, in which case photon absorption is not possible. Indeed, in this case, Eq.~\eqref{eq:height_function} implies $\#\{g_m|\verb|l|(g_m)=j\}=0$, or in other words, a suitable $g_a$ does not exist. In order to absorb the next photon, we must therefore first find a way to increase $h(j)$ relative to $h(j-1)$. This can be accomplished with a time-reversed measurement on an emitter. To perform this operation, we first rotate the state to $\Ket{\Phi}\otimes\Ket{0}_i$, where $\Ket{\Phi}$ is a stabilizer state involving photons $1,\ldots,j$ and emitters other than $i$. This can always be done using $\mathcal{O}(n_e)$ Clifford gates on emitters when $h(j)<h(j-1)$ (see Methods). 
Now notice that this state is obtained from the pre-measurement state CNOT$_{ij}\Ket{\Phi}\otimes\Ket{+}_i$ when emitter $i$ is measured to be in the state $\Ket{0}_i$. Therefore, starting from $\Ket{\Phi}\otimes\Ket{0}_i$, if we perform a Hadamard gate on emitter $i$ followed by the gate CNOT$_{ij}$, we effectively reverse the measurement on the emitter. These operations transform the stabilizers $g_m$ in such a way that $h(j)$ now satisfies $h(j)\ge h(j-1)$ (see Methods), and we can proceed with the next photon absorption. The emitter gates, Hadamard on $i$, and CNOT$_{ij}$
are all included in the time-reversed generation sequence.

\subsection{Examples}\label{sec:examples}

We demonstrate our algorithm with several examples. The first is the important case of repeater graph states \cite{Azuma2015}, where we use our algorithm to obtain generation protocols that are more efficient than previously known ones. As a second example, we consider random graphs containing up to hundreds of photons and demonstrate the polynomial scaling of the resulting generation circuits. Additional examples, including modified repeater graph states, error correcting codes, and a simple example that illustrates the algorithm in detail can be found in Supplementary Notes 1-4.

Next, we apply our algorithm to find operation sequences that produce repeater graph states \cite{Azuma2015}. In addition to its importance in quantum network applications, this example also illustrates how different photon emission orderings impact the required number of emitters. Ref.~\cite{Buterakos_2017PRX} presented a generation protocol for a particular ordering that was devised essentially through guesswork. Our algorithm can be used to systematically find protocols for any ordering. An example of a 12-photon repeater graph state is shown in Fig.~\ref{fig:RGSs}(a). The graph contains a fully connected core of 6 photons, each of which is connected to a single external photon. Bell measurements are performed on pairs of these external photons, where the two photons in each pair come from different graph states. If a Bell measurement succeeds, then the two corresponding core photons are linked by an edge, and entanglement extends across two nodes of the repeater network. Having multiple external photons provides built-in redundancy that increases the likelihood that at least one Bell measurement between two repeater graph states is successful. Upon success, core photons are then measured in the $z$ or $x$ basis to remove photons connected to failed measurements or to create entanglement links between successful measurements, respectively. Because the external photons are measured first, it may be advantageous to emit these first when generating the graph state to reduce photon storage requirements. This corresponds to the photon ordering shown in Fig.~\ref{fig:RGSs}(a). The height function for this graph and photon ordering is shown in Fig.~\ref{fig:RGSs}(d), where it is evident that 6 emitters are needed to produce the state. However, if efficient photon storage is available, then the ordering shown in Fig.~\ref{fig:RGSs}(b) may be preferable, where now external and core photons are emitted in an alternating sequence. This ordering reduces the number of emitters down to only 2, as shown in Fig.~\ref{fig:RGSs}(e). As we discuss further below, this illustrates our general finding that ``natural" orderings in which neighboring vertices are emitted around the same time reduce the requisite number of emitters. This reduction in quantum resources becomes still more dramatic as the size of the graph increases; for orderings as shown in Fig.~\ref{fig:RGSs}(a), the number of emitters scales linearly with photon number, while for the natural ordering of Fig.~\ref{fig:RGSs}(b), the number of emitters remains at 2 regardless of the number of photons. This is shown explicitly in the Supplemental Information.

As discussed in Ref.~\cite{Russo_2019}, some of the edges in the repeater graph can be removed without affecting the functionality of the repeater. Fig.~\ref{fig:RGSs}(c) shows an example of this in which 4 of the core edges are deleted. As shown in Fig.~\ref{fig:RGSs}(f), the number of emitters is still 2. However, removing the redundant edges reduces the depth of the resulting generation circuit, which is shown in Fig.~\ref{fig:RGSs}(g). This circuit contains 4 CNOTs between emitters and 1 intermediate measurement on an emitter, whereas the original protocol presented in Ref.~\cite{Buterakos_2017PRX} requires 5 two-qubit gates and 5 intermediate measurements.

\begin{figure}[h]
	\centering
	\includegraphics[width=0.9\linewidth]{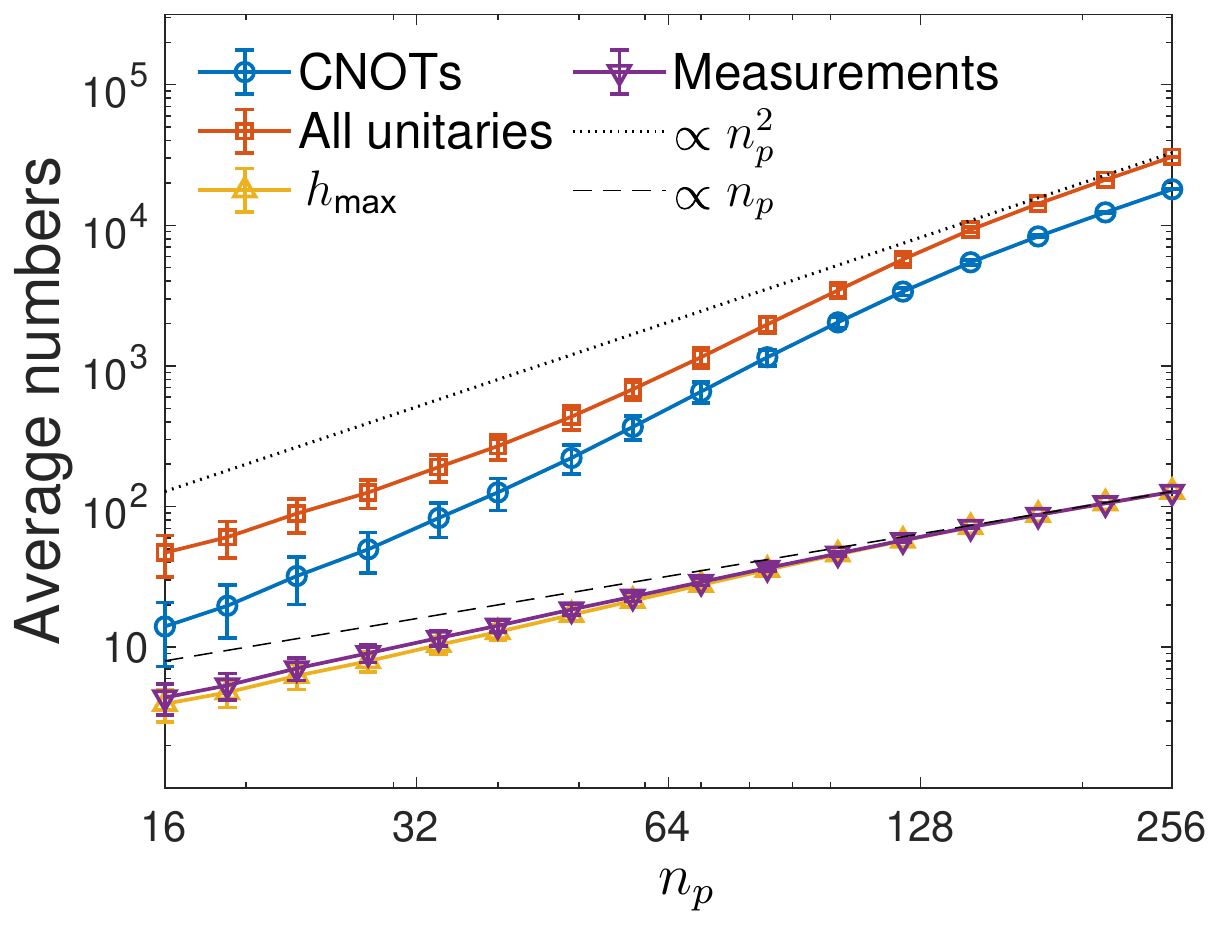}
	\caption{\textbf{Scaling of emitter number and generation circuit depth.} The maximum value of the height function $h_{\text{{max}}}$, measurement counts, and gate counts needed to produce random graphs of size $n_p$ are all averaged over 1024 graph realizations for each value of $n_p$ drawn from an Erd\"os–R\'enyi ensemble with edge probability $p = 0.95$. Dashed curves are included to show the scaling with $n_p$ and $n_p^2$. The error bars stand for the standard deviation of these realizations.
	}
	\label{fig:random}
\end{figure}

To demonstrate how our algorithm scales with the number of photons in the target state, we run it for random graphs ranging in size from $n_p=16$ to $n_p=256$ photons. These graphs are produced randomly using the Erd\"os–R\'enyi model \cite{RandomGraphs}. In this approach, each random graph is constructed by connecting $n_p$ vertices randomly with fixed probability $p$. We discard any graphs that contain disconnected vertices when sampling these realizations. The likelihood that such graphs arise becomes very small if $p$ is chosen sufficiently close to 1. In Fig.~\ref{fig:random}, we show the maximum value, $h_{\text{max}}$, of the height function averaged over 1024 realizations for each value of $n_p$. Averaged measurement and gate counts are also shown. It is evident that $h_{\text{max}}$, and hence the number of emitters, scales linearly with $n_p$ as $n_p$ becomes large. The same is also true of the number of measurements. On the other hand, the number of CNOTs and the total number of gates in the resulting generation circuits scale quadratically with the number of photons in the target state. These results confirm both the polynomial scaling of our algorithm, which allows us to easily find generation protocols for graph states containing hundreds of photons, and the polynomial scaling of the resulting protocols, which makes them practical for near-term experiments.

\subsection{Photon emission ordering}\label{sec:discussion}

A powerful feature of our algorithm is that it readily incorporates a desired photon emission ordering. This is encoded when we arrange the photons and emitters in a 1D lattice to define the height function. If no specific ordering is preferred, then ideally we would want to choose the ordering that minimizes the number of emitters $n_e$. However, the task of finding this optimal ordering is NP-hard, as we show in Methods. Nevertheless, one can still look for heuristic solutions to the problem. In fact, the expression for the height function in Eq.~\eqref{eq:height_function} makes it clear that this function is suppressed for orderings in which the stabilizers, when expressed in the echelon gauge, are supported predominantly on high-index sites on the right side of the 1D lattice. This tends to occur for ``natural" orderings in which neighboring photons in the graph are emitted around the same time, because in this case the stabilizers are localized on the 1D lattice. This was illustrated with our repeater graph state example in the previous section. The extent to which the stabilizers can be localized in this way depends on the graph of course. For an $N\times M$ square lattice, it is inevitable that some neighboring vertices will be separated by $M$ steps in the emission sequence (assuming $M<N$), and so the number of emitters is of order $M$. On the other hand, for other graph structures like those of the repeater graph states, far fewer emitters may be needed, provided a natural photon ordering is used. Note that in this example, as for many graphs, edges between remote vertices cannot be avoided (see Fig.~\ref{fig:RGSs}(b)). Despite this, we showed that optimal orderings for which the height function remains small can still be found. Thus, emitting neighboring vertices around the same time is sufficient but not always necessary to keep the number of emitters small.

In summary, we presented an efficient algorithm to construct polynomial-depth operation sequences that produce arbitrary multi-photon graph states from a minimal number of quantum emitters. By reducing both the number of photon sources and the number of quantum operations that need to be performed on them, our method brings the wide range of quantum information applications that rely on entangled photon resource states closer to experimental reality.

\section{Methods}\label{sec:methods}

\subsection{Echelon gauge}

The echelon gauge was first defined in Ref.~\cite{Audenaert_2005NJP}, where it was called row reduced echelon form. In this gauge, the stabilizer tableau has a recursive row-reduced form based on the following three types of matrices:
\begin{equation}\label{eq:echelon_gauge}
	\left(\begin{array}{l|cr} \mathbbm{1} & *\cdots * & \\ \hline \mathbbm{1} & & \\ \vdots & M & \\ \mathbbm{1} & & \end{array}\right),\quad \left(\begin{array}{l|cr} \sigma & *\cdots * & \\ \hline \mathbbm{1} & & \\ \vdots & M & \\ \mathbbm{1} & & \end{array}\right), \quad \left(\begin{array}{l|cr} \sigma_1 & *\cdots * & \\ \sigma_2 & *\cdots * & \\ \hline \mathbbm{1} & & \\ \vdots & M & \\ \mathbbm{1} & & \end{array}\right),
\end{equation}
where $\sigma$, $\sigma_1$, and $\sigma_2$ are nontrivial Pauli matrices, and $\sigma_1\ne\sigma_2$. In this work, we always choose $\sigma_2 = \sigma^z$, and $\sigma_1$ can be either $\sigma^x$ or $\sigma^y$. The full tableau cannot have the first form shown above (with only identities in the first column), because this case does not apply to pure states. However, the submatrix $M$ can follow any of the above three patterns, and the structure iterates recursively. The stabilizers can be transformed into this gauge starting from any other by performing a series of row reductions, as described in Ref.~\cite{Audenaert_2005NJP}. In the echelon gauge, the independent stabilizers acting on $\bar{A}=\{x+1,\ldots,n\}$ appear at the bottom right of the tableau. Therefore, starting from the formula for the entanglement entropy for subregion $\bar{A}$ of a stabilizer state \cite{Fattal2004}, $S_{\bar{A}} = n_{\bar{A}} - |\mathcal{G}_{\bar{A}}|$, where $n_{\bar{A}}$ is the size of $\bar{A}$ and $|\mathcal{G}_{\bar{A}}|$ is the number of independent stabilizers acting on $\bar{A}$, and using $h(x)=S_A = S_{\bar{A}}$, we obtain Eq.~\eqref{eq:height_function}.

\subsection{Time-reversed measurements}

Above, we saw that when the total state of the system has the form $\Ket{\Phi}\otimes\Ket{0}_i$, where $i$ is an emitter site, we can perform a time-reversed measurement to convert this to the pre-measurement state CNOT$_{ij}\Ket{\Phi}\otimes\Ket{+}_i$. Here, we clarify two important questions regarding this process: (i) When and how can we bring the system into the state $\Ket{\Phi}\otimes\Ket{0}_i$? (ii) How can we see that a time-reversed measurement on this state increases $h(j)$, as needed for a subsequent photon absorption process?

Regarding question (i), when $h(j)<h(j-1)$, we can always find a set of Clifford gates that act purely on the emitters that will transform the state of the system into $\Ket{\Phi}\otimes\Ket{0}_i$. To see this, first note that $h(j)=h(n_p)$, as follows from Eq.~\eqref{eq:height_function} when photons $j+1$ through $n_p$ are in state $\Ket{0}$. Using that the height function is bounded from above by $n_e$, we then have $h(n_p)=h(j)<h(j-1)\le n_e$. On the other hand, from Eq.~\eqref{eq:height_function} we have $h(n_p)=n_e-\#\{g_m|\verb|l|(g_m)>n_p\}$. Together, these results imply $\#\{g_m|\verb|l|(g_m)>n_p\}>0$, or in other words, there is at least one stabilizer that is supported solely on the emitter sites. We can therefore transform this stabilizer into $\sigma_i^z$ using at most $\mathcal{O}(n_e)$ Clifford gates on the emitters, bringing the state to $\Ket{\Phi}\otimes\Ket{0}_i$. We can then convert this stabilizer to $\sigma_i^x$ by applying a Hadamard gate on site $i$. This prepares the system for the second part of the time-reversed measurement process, which is the gate CNOT$_{ij}$.

We answer question (ii) by proving the following theorem:

 \begin{theorem}\label{theorem}
	Theorem 1: If $h(j) < h(j-1)$ and the $i$-th qubit $(i>j)$ is stabilized by $\sigma^x_i$, then applying $\mathrm{CNOT}_{ij}$ will boost $h(x)\to h(x)+1$, $\forall x\in \{j,j+1,\cdots,i-1\}$.
	 \end{theorem}

~

\paragraph*{Proof}
We are assuming that $h(j)<h(j-1)$, which from Eq.~\eqref{eq:height_function} implies $\#\{g_m|\verb|l|(g_m)=j\}=0$. Now consider how the stabilizers transform under $\mathrm{CNOT}_{ij}$. If $\verb|l|(g_m)<j$ before the gate, then $\verb|l|(g_m)$ remains invariant, and the contributions of these stabilizers to $h(x)$ remain the same after the gate. The only potential changes to $h(x)$ come from stabilizers $g_m$ for which $\verb|l|(g_m)>j$. These stabilizers necessarily have $\mathbbm{1}$ on the $j$-th site. Stabilizers among this set that have $\mathbbm{1}$ or $\sigma_i^z$ on the $i$-th site will be unchanged by the $\mathrm{CNOT}_{ij}$ gate. However, if one or more of these stabilizers have $\sigma_i^x$ or $\sigma_i^y$ before the gate, then afterward, these stabilizers will contain $\sigma_j^x$. Consequently, $h(j)$ increases, while $h(j-1)$ remains the same. In the echelon gauge, there can only be one stabilizer with $\sigma_j^x$ as the left-most nontrivial Pauli. Therefore, $h(j)\to h(j)+1$ when $\mathrm{CNOT}_{ij}$ is applied. Moreover, if the $i$-th qubit is stabilized by $\sigma_i^x$, then this becomes $\sigma_j^x\sigma_i^x$ after the gate, and so the height function for all sites between $j-1$ and $i$ increases: $h(x)\to h(x)+1$ $\forall x\in \{j,j+1,\cdots,i-1\}$. $\Box$

\subsection{Scaling analyses}

Here, we determine the complexity of both the protocol solver algorithm itself and the resulting graph state generation circuit. Regarding the algorithm, the main factor that determines the complexity is the need to restore the stabilizers to the echelon gauge after each operation is applied. Transforming a $n$-qubit stabilizer state into the echelon gauge generally requires $\mathcal{O}(n^3)$ steps, which is the complexity of Gaussian elimination. Another important factor is the process of determining which gates need to be applied in preparation for photon absorption or time-reversed measurement. Solving for each set of gates takes no more than $\mathcal{O}(n_e n)$ steps, which is the number of entries in the emitter part of the stabilizer tableau. Thus, the Gaussian eliminations needed to restore echelon gauge dominate the scaling. In the worst case where $n_e\propto n$, our algorithm will then take $\mathcal{O}(n^4)$ steps, where the additional factor of $n$ comes from the fact that the algorithm requires $\mathcal{O}(n_p)\sim\mathcal{O}(n)$ iterations.

As for the complexity of the output generation circuit, there are at most $\mathcal{O}(n_e)$ operations between any two-photon emissions. For example, $\mathcal{O}(n_e)$ gates are needed to transform $g_a$ into the appropriate form for photon absorption. Thus, the depth of the circuit acting on the emitter qubits is at most $\mathcal{O}(n_p n_e)$. In the worst case where $n_e\sim n_p$, the scaling is then $\mathcal{O}(n_p^2)$, which is consistent with Fig.~\ref{fig:random}.
Nevertheless, due to the fact that some long-range two-qubit gates may arise, and given that these are usually decomposed as $\mathcal{O}(n_e)$ short-ranged two-qubit gates in real devices, the overall circuit depth may become $\mathcal{O}(n_p n_e^2)$.

\subsection{Complexity of finding optimal photon emission orderings}\label{sec:optimal_orderings}

We can show that the task of finding optimal emission orderings is NP-hard by mapping this to a known graph theory problem. Define $\Gamma_{ij}$ to be the adjacency matrix of the graph representing the target state $\Ket{\Psi}$. Ref. \cite{Hein2004PRA} showed that we can obtain the height function from $\Gamma_{ij}$ using the formula  $h(x)=\mathrm{rank}_{2}(\Gamma_{A\bar{A}})$, where $\Gamma_{A\bar{A}}$ is the sub-matrix of $\Gamma_{ij}$ with row indices $i\in A=\{1,2,\cdots,x\}$ and column indices $j\in \bar{A}$.
Note that this expression does not simplify the computation of $h(x)$; it can take more steps to find the maximum compared to using Eq.~\eqref{eq:height_function} since the former performs Gaussian eliminations for $\mathcal{O}(n_p)$ rounds, while the latter only takes one round.
However, the optimized maximum value of this alternative expression for $h(x)$ (i.e., $\mathrm{max}_xh(x)$) is precisely equal to a graph theoretic property known as linear rank-width (LRW) \cite{OUM201715}. The task of finding an optimal photon emission ordering is therefore equivalent to finding LRW through the graph isomorphism, which has long been studied in coding theory in the context of optimizing block code trellises \cite{Massey1978}.
Unfortunately, determining whether a simple connected graph has an LRW bounded from above by a positive integer $k$ has been shown to be NP-complete \cite{OUM200579,Jeong2017}. Therefore, it is unlikely this problem can be solved efficiently for large, arbitrary photonic graph states unless $\text{P}=\text{NP}$.
Nevertheless, if the parameter $k$ is set to $1$, this problem can be answered in polynomial time \cite{Adler2017}. If the parameter $k$ is set to larger values, a recent work \cite{Jeong2017} showed that this problem can be reduced to a fixed parameter tractable problem. Specifically, its answer, along with the sequence solution (if it exists), can be determined in $\mathcal{O}(f(k)n^3_p)$ steps, where $f(k)$ is an exponentially large function of $k$. However, the growth of $f(k)$ is so rapid that this result is not likely to be of practical use for photonic graph state generation.

\section*{Data Availability}
The data that support the findings of this study are available from the authors upon
request.

\section*{Code Availability}
A custom MATLAB code to reproduce our results is available on GitHub and archived in Zenodo (https://doi.org/10.5281/zenodo.5652105).

\section*{Acknowledgments}

This work was in part supported by the National Science Foundation (grant no. 1741656). E.B. acknowledges support by National Science Foundation grant no. 2137953. S.E.E. acknowledges support by the Army Research Office (MURI grant no. W911NF2120214).

\section*{Author Contributions}
E.B and S.E.E. conceived and supervised the project. B.L. developed the general approach and performed all the calculations. E.B. and S.E.E. contributed to technical components of the project. All authors contributed to the writing of the manuscript.

\section*{Competing Interests}
The authors declare that there are no competing interests.

\nocite{}
\bibliographystyle{unsrt}


\providecommand{\noopsort}[1]{}\providecommand{\singleletter}[1]{#1}%

	\newpage
	\newpage
	\widetext
	\begin{center}
		\LARGE{\textbf{Supplementary Information} }
	\end{center}
	\setcounter{section}{0}
	\setcounter{equation}{0}
	\setcounter{figure}{0}
	\setcounter{table}{0}
	\makeatletter
	\renewcommand{\theequation}{S\arabic{equation}}
	\renewcommand{\thefigure}{S\arabic{figure}}
	\renewcommand{\bibnumfmt}[1]{[S#1]}
	\renewcommand{\citenumfont}[1]{S#1}

\section*{Supplementary Note 1}
This supplementary material contains several additional examples of generation protocols produced using our algorithm. We begin with a simple example that illustrates in detail how our algorithm works. We then provide solutions for more complicated examples of practical importance, including error correcting codes and repeater graph states of arbitrary size. These more complicated examples are solved numerically using MATLAB codes that are available on GitHub \cite{MATLAB_CircuitSolver}. This code expresses the generation sequence in terms of an MPS \cite{Schon2005PRL_S} for bookkeeping purposes:
\begin{equation}\label{eq:MPSsolution}
	\Ket{\Psi}
	=
	U_{p,\mathrm{tot.}}\langle\psi_f|\big[\prod_{j = 1}^{n_p}\left(\hat{M}_{j}U_{e,j}\hat{E}_{\eta_{j}}\right)\big]W_0 |\psi_0\rangle,
\end{equation}
in which the initial and final states of emitter qubits are simply product states of $\Ket{0}$: $\Ket{\psi_f} = \Ket{\psi_0} = \Ket{0}^{\otimes n_e}$. 
We denote $\eta_j$ as the emitter qubit that emits the $j$-th photon and $\mu_j$ as the emitter qubit that is measured after emitting the $j$-th photon.
$\hat{E}_{\eta_j}$ is the emission tensor, 
$\hat{E}_{\eta_j} = \Ket{0}_{j}  \Ket{0}_{\eta_j}\!\!\Bra{0}_{\eta_j} + \Ket{1}_{j}  \Ket{1}_{\eta_j}\!\!\Bra{1}_{\eta_j}$,
that describes emission of photon $j$ from emitter $\eta_j$, which can be represented as $\mathrm{CNOT}_{\eta_j,j}$.  
$U_{e,j}$ is the unitary operation obtained from the $j$-th photon absorption step, which transforms $g_a$ as explained in the main text (Sec.II.A).
$\hat{M}_j$ is identity if no measurement happens ($\mu_j$ is not assigned), otherwise, $\hat{M}_j = 
W_{j}H_{\mu_j}X^{s_j}_{\mu_j}\hat{\pi}_{\mu_j}$, with projection $\hat{\pi}_{\mu_j} \equiv \frac{1}{2}[\mathbbm{1}+(-1)^{s_j}Z_{\mu_j}]$ and its random outcome $s_j\in\{0,1\}$. 
Here, $W_j$ is the unitary operation that is obtained from time-reversed measurement (Sec.IV.B), and $W_0$ is the unitary operation that disentangles all emitters at the final stage of the time-reversed sequence.
Finally, $U_{p,\mathrm{tot}} = \prod_j\left(X^{s_j}_j U_{p,j}\right)$ is the local Clifford operation that acts on photons with conditional $X^{s_j}_j$ flipping. The profile of the solution is stored as $\{U_{e,j},U_{p,j},\mu_j,\eta_j,W_j,W_0\}$. We note that such a solution is usually not unique due to there being multiple choices for how to choose the emitter gates and emitter sites in each photon absorption and time-reversed measurement.

As discussed in the main text, the height function plays a central role in determining the number of emitters and the operation sequence needed to generate a target photonic graph state. As shown in Eq.~(1) of the main text, when the stabilizers $g_m$ are in the echelon gauge, the height function can be expressed as
\begin{equation}\label{suppeq:height_function}
	h(x)=n - x - \#\{g_m|\verb|l|(g_m)>x\},
\end{equation}
where $\verb|l|(g_m)$ is the index of the left-most (smallest index) site on which $g_m$ acts nontrivially. In the main text, we showed that the difference in the height function across adjacent sites determines whether we perform a photon absorption or a time-reversed measurement at each step of the algorithm. Therefore, we define
\begin{equation}\label{eq:deltah}
	\delta h(x)\equiv h(x)-h(x-1) = \#\{g_m|\verb|l|(g_m)=x\}-1,
\end{equation}
from which it is apparent that this difference only depends on the number of stabilizers (in the echelon gauge) that have a left-ending on site $x$.

We begin by demonstrating our protocol solver algorithm in the case of the simple 4-photon graph state displayed in Supplementary Fig.~\hyperref[fig:SlashSquareSolving]{\ref*{fig:SlashSquareSolving}}(a). The stabilizers are given by
\begin{equation}\label{eq:g1g2g3g4}
	\begin{aligned}
		g_1 &= \sigma^x_1\sigma^z_2\sigma^z_3,\quad
		g_2 = \sigma^z_1\sigma^x_2\sigma^z_3\sigma^z_4,\\
		g_3 &= \sigma^z_1\sigma^z_2\sigma^x_3\sigma^z_4,\quad
		g_4 = \sigma^z_2\sigma^z_3\sigma^x_4.\\
	\end{aligned}
\end{equation}
We can switch to the echelon gauge by redefining $g_3\to g_2g_3$. We then calculate the height function using Supplementary Eq.~\eqref{suppeq:height_function}, finding that the maximum is $2$. Therefore, at least $n_e=2$ emitter qubits are needed, and so we assemble a $6$-qubit lattice. We can depict the complete set of 6 stabilizers as a tableau, as shown in Supplementary Fig.~\hyperref[fig:SlashSquareSolving]{\ref*{fig:SlashSquareSolving}}(b).

In Supplementary Fig.~\hyperref[fig:SlashSquareSolving]{\ref*{fig:SlashSquareSolving}}(c), we first obtain inset (1) by transforming Supplementary Fig.~\hyperref[fig:SlashSquareSolving]{\ref*{fig:SlashSquareSolving}}(b) to the echelon gauge. The upper left sub-block of the tableau is exactly Supplementary Eq.~\eqref{eq:g1g2g3g4} with $g_3\to g_2g_3$. Next we describe in detail how the generator set is updated from inset (1) to inset (17) step by step. The column label $j$ indicates which photon we are currently focusing on, and the labels (i),...,(iv) indicate the specific step of our algorithm. For each photon, we do the following steps:
\begin{itemize}
	\item $j=4$: (i) Obtain inset (1) by transforming to echelon gauge: $g_3\to g_2g_3$. 
	(ii) Supplementary Eq.~\eqref{eq:deltah} gives $\delta h(4)=-1$. Perform a time-reversed measurement on emitter site 5 by applying a Hadamard $H_5$ followed by $\mathrm{CNOT}_{54}$, which yields inset (2). 
	(iii) Let $g_a = g_5 = \sigma^x_4\sigma^x_5$ in inset (2). One gets inset (3) by performing Hadamards on sites 4 and 5. Then the $4$-th photon is absorbed into the emitter on site 5 by applying $\mathrm{CNOT}_{54}$. Replace $g_4\to g_4g_5$ to eliminate the redundant $\sigma^z_4$, yielding inset (4). ($\mu_4=5$, $\eta_4=5$, $U_{p,4}=H_4$, $U_{e,4} = H_5$, $W_4=H_5$.)
	
	\item $j=3$: (i) Skip this step since inset (4) is already in echelon gauge. 
	(ii) Supplementary Eq.~\eqref{eq:deltah} gives $\delta h(3) =-1$. Perform a time-reversed measurement on emitter site 6 by applying $H_6$ followed by $\mathrm{CNOT}_{63}$. Inset (6) is then obtained by redefining $g_5\leftrightarrow g_6$. 
	(iii) Let $g_a = g_5 = \sigma^x_3\sigma^x_6$ in inset (6). One gets inset (7) by applying Hadamards on sites 3 and 6. Then the $3$rd photon is absorbed by applying $\mathrm{CNOT}_{63}$. Replace $g_3\to g_3g_5$ to eliminate the redundant $\sigma^z_3$. Thus, inset (7) becomes (8). ($\mu_j=6$, $\eta_3=6$, $U_{p,3}=H_3$, $U_{e,3} = H_6$, $W_3=H_6$.)
	
	\item $j=2$: (i) Skip this step since inset (8) is already in echelon gauge. 
	(ii) Skip this step since Supplementary Eq.~\eqref{eq:deltah} gives $\delta h(2)=1$. 
	(iii) Choose $g_a = g_4 = \sigma^z_2\sigma^z_5\sigma^x_6$ in inset (10). One gets inset (11) by applying  $H_6$ followed by $\mathrm{CNOT}_{65}$ on the emitters, so that $g_a\to\sigma^z_2\sigma^z_5$. Then the $2$nd photon is absorbed into emitter 5 by applying $\mathrm{CNOT}_{52}$. Redefine $g_k\to g_kg_4$ for $k=1,3$ to eliminate the redundant $\sigma^z$'s. Thus, inset (11) becomes (12). 
	($\eta_2=5$, $U_{p,2}=\mathbbm{1}$, $U_{e,2} = H_6\mathrm{CNOT}_{65}$, $\hat{M}_2 = \mathbbm{1}$.)
	
	\item $j=1$: (i) Obtain inset (13) from (12) by transforming to echelon gauge: $(g_3,g_4,g_5,g_6)\rightarrow (g_4,g_5,g_6,g_3)$. 
	(ii) Skip this step since Supplementary Eq.~\eqref{eq:deltah} gives $\delta h(1)=1$. 
	(iii) Choose $g_a = g_2 = \sigma^z_1\sigma^x_5\sigma^z_6$ in inset (14). One gets inset (15) by applying $H_5$ and then $\mathrm{CNOT}_{65}$ to transform $g_a\to\sigma^z_1\sigma^z_5$. Then the $1$st photon is absorbed into emitter site 5 by applying $\mathrm{CNOT}_{51}$. Thus, inset (15) becomes (16).
	($\eta_1=5$, $U_{p,1}=\mathbbm{1}$, $U_{e,1} = H_5\mathrm{CNOT}_{65}$, $\hat{M}_1 = \mathbbm{1}$.)
	
	\item (iv) Finally, to recover the state $\Ket{0}^{\otimes n}$, one needs to disentangle the emitter qubits. This can be done with the following gate sequence: $H_5\mathrm{CNOT}_{56}H_5$. In the last step, we permute the $g_m$ to obtain inset (17). ($W_0=H_6\mathrm{CNOT}_{56}H_5$.)
\end{itemize}
\begin{figure}[t]
	\centering
	\includegraphics[width=1.0\linewidth]{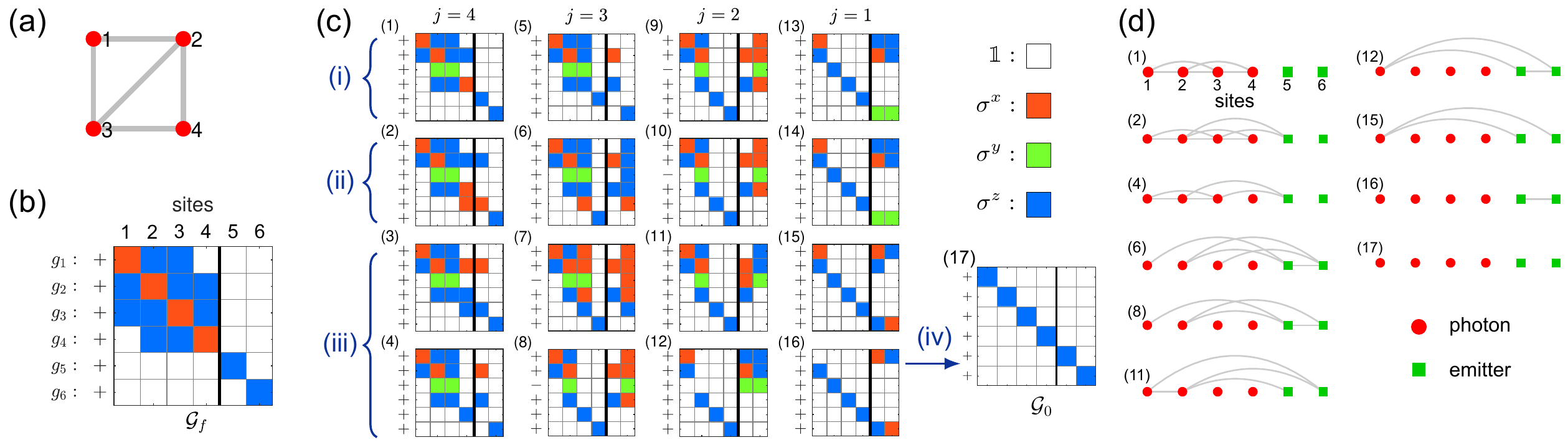}
	\caption{\textbf{Step-by-step illustration of the protocol solver.} (a) A target graph state with 4 photons. (b) The set of generators $\mathcal{G}_f =\{g_m\}$ is depicted as a tableau in which each row corresponds to one generator. Different colors correspond to different Pauli operators. The first 4 columns correspond to photonic qubits, and the last 2 columns correspond to emitters. (c) Step by step demonstration of how to obtain the time-reversed generation sequence, where $\mathcal{G}_0=\{\sigma^z_i\}$ is finally obtained. Explanations are in the main text.
		(d) Local Clifford equivalent graph state representations of tableaux in (c).}
	\label{fig:SlashSquareSolving}
\end{figure}
\begin{figure}[h]
	\centering
	\includegraphics[width=1\linewidth]{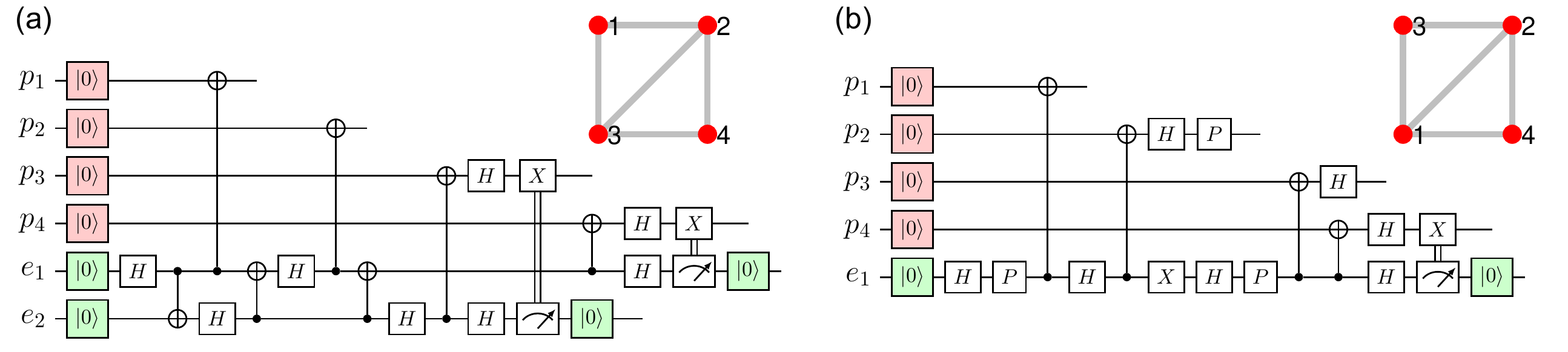}
	\caption{
		\textbf{Graph state generation circuits.}
		In this figure, $p_j \;(j=1,2,3,4)$ labels different photonic qubits, and $e_1$ and $e_2$ are emitter qubits. 
		At the end of each circuit, the photon qubits are in the target graph state displayed at the top right, while the emitter qubits are in state $\Ket{0}$ after the measurements.
		(a) The emission circuit obtained from the steps in Supplementary Fig.~\hyperref[fig:SlashSquareSolving]{\ref*{fig:SlashSquareSolving}}. 
		(b) A different generation circuit that produces the same target graph state as in (a). This circuit is obtained by swapping qubits $1\leftrightarrow 3$, resulting in a circuit that requires only one emitter.
	}
	\label{fig:SlashSquareCircuit}
\end{figure}
Now that the algorithm is complete, we reverse all the operations to obtain the final generation sequence. This circuit is shown in Supplementary Fig.~\hyperref[fig:SlashSquareCircuit]{\ref*{fig:SlashSquareCircuit}}(a). 
It is worth noting that, in this example, the emission sequence can be further optimized by swapping the 1st and 3rd photons in the emission order, such that the maximum of $h(x)$ is reduced to $1$. Thus, only one emitter qubit is needed in this case, and the corresponding generation circuit is displayed in Supplementary Fig.~\hyperref[fig:SlashSquareCircuit]{\ref*{fig:SlashSquareCircuit}}(b).

\section*{Supplementary Note 2}
In this subsection, we demonstrate how to generate a useful quantum error correction code, with some continuous logical rotation. In particular, we present an emission sequence for the Shor code \cite{ShorCode1995_S} with 9 photonic qubits, which is able to protect a qubit from single bit-flip and phase-flip errors. The stabilizer generators of this code are well known:
$g_{j} = \sigma^z_j\sigma^z_{j+1}$ for $j = 1,2,4,5,7,8$, and $g_3 = \sigma^x_1\sigma^x_2\cdots \sigma^x_6$, $g_6 = \sigma^x_4\sigma^x_5\cdots \sigma^x_9$ \cite{nielsen_chuang_2010_S}. 
We can also define the logical operators $X_L \equiv \sigma^z_1\sigma^z_2\cdots \sigma^z_9$, $Z_L \equiv \sigma^x_1\sigma^x_2\cdots \sigma^x_9$ and $Y_L = iX_LZ_L$. 
Let the last stabilizer be $g_9 = \pm X_L$, which determines a pair of logical space basis states $\Ket{\pm}_L$. For both choices, Supplementary Eq.~\eqref{suppeq:height_function} gives $h_{\max} = 2$, and the emission circuit solutions for $\Ket{\pm}_L$ are given in Supplementary Fig.~\hyperref[fig:ShorCircuit]{\ref*{fig:ShorCircuit}}, where $\Ket{\pm}_L$ are separately given by $R = \mathbbm{1}$ and $R = X_{e_1}$. Therefore, by replacing $R$ by a more general $x$-rotation, $e^{i\frac{\varphi}{2}X_{e_1}} \equiv \mathbbm{1}\cos\frac{\varphi}{2} + i X_{e_1}\sin\frac{\varphi}{2}$, we can obtain a rotated logical qubit $\Ket{\varphi}_L = e^{i\frac{\varphi}{2}Z_L}\Ket{+}_L$, with an arbitrary angle $\varphi$.
That is, the circuit in Supplementary Fig.~\hyperref[fig:ShorCircuit]{\ref*{fig:ShorCircuit}} allows us to transmit a rotated photonic logical qubit protected by the Shor code, 
with merely $2$ emitter qubits, $1$ two-qubit gate and $2$ measurements, which is surprisingly simple.
\begin{figure}[h]
	\centering
	\includegraphics[width=0.5\linewidth]{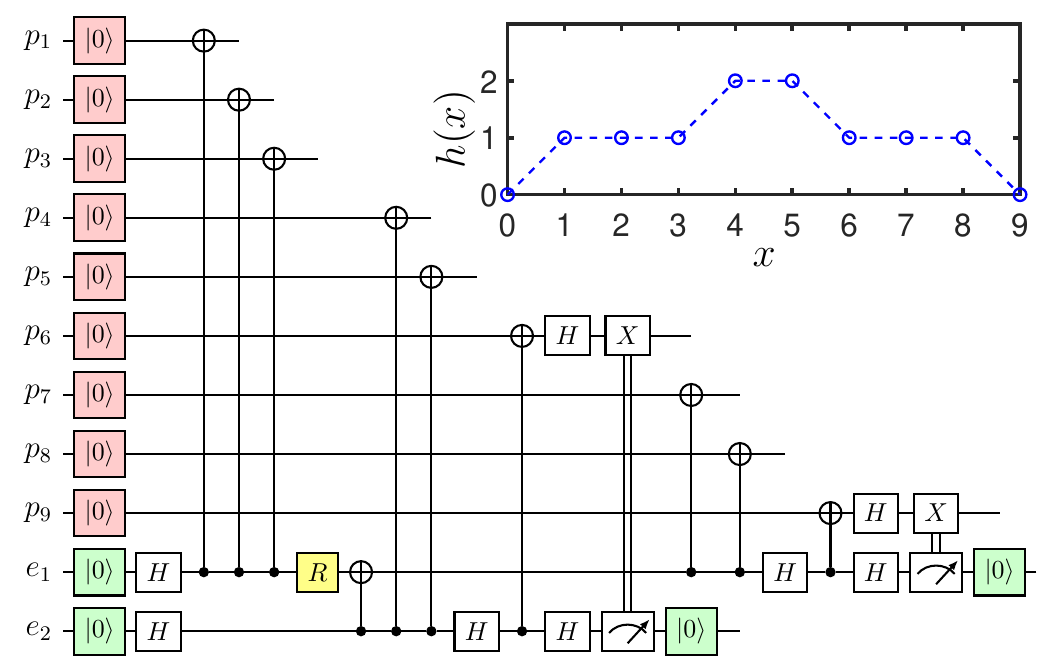}
	\caption{\textbf{Shor code example.} The emission circuit that generates a logical state of the Shor code, controlled by a local operation $R$ (the yellow block). The inset displays the height function of $\Ket{\pm}_L$, which has $h_{\max} = 2$. 
	}
	\label{fig:ShorCircuit}
\end{figure}

\section*{Supplementary Note 3}
In this section, we generalize the repeater graph state example from the main text and present explicit generation circuits for repeater graphs with arbitrarily many photons. As shown in Supplementary Fig.~\hyperref[fig:RGS2m]{\ref*{fig:RGS2m}}, for a repeater graph state with $2m$ photons, the maximum of the height function indicates that we need $2$ emitters regardless how large $m$ is $(m\ge4)$. The unitary operations $A$, $B$ and $C$ displayed in Supplementary Fig.~\hyperref[fig:RGS6m]{\ref*{fig:RGS6m}}(a) depend on $m$:
\begin{equation}
	A = X_{e_2}^{\lfloor m/2\rfloor+1} ,\quad B = X_{e_1}^{\lfloor m/2\rfloor}, \quad C = P^{m}_{e_2} \;,
\end{equation}
where $X_i=\sigma_i^x$, and $P=\mathrm{diag}(1,i)$.
Compared to the approach given in previous work \cite{Buterakos_2017PRX}, which requires $m-1$ two-qubit gates and $m$ measurements, the new solution in Supplementary Fig.~\hyperref[fig:RGS2m]{\ref*{fig:RGS2m}}(a) uses $2m - 3$ two-qubit gates and $m-1$ measurements, reducing the number of measurements needed to produce the state. We highlight that our method yields different solutions with flexible settings, so actually the solution from Ref.~\cite{Buterakos_2017PRX_S} can also be obtained from our algorithm. 
\begin{figure}[h]
	\centering
	\includegraphics[width=1.0\linewidth]{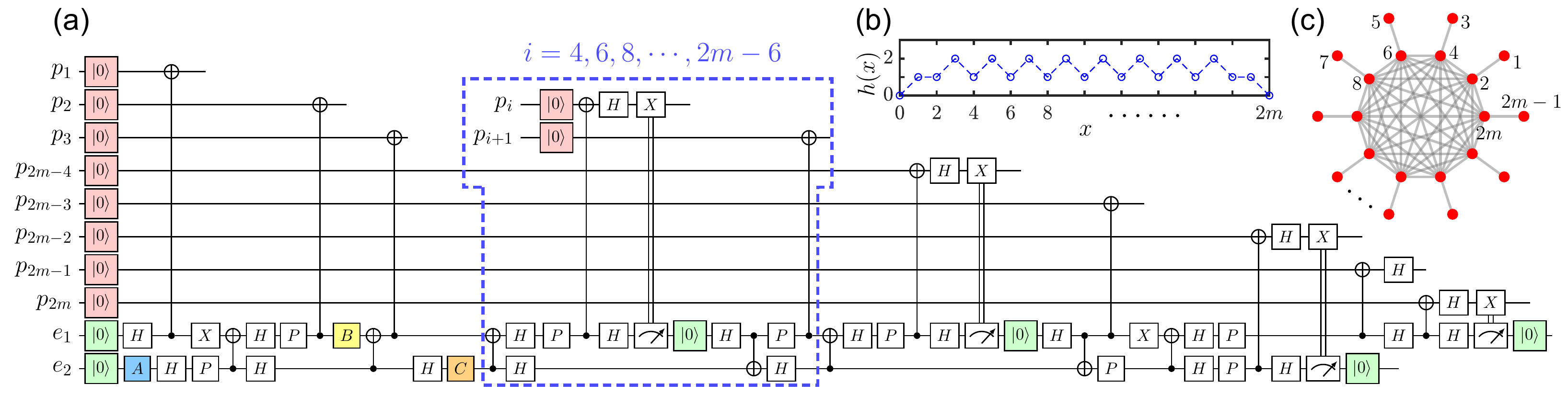}
	\caption{
		\textbf{Example of a large RGS.}
		(a) and (b) show the emission circuit and height function for the repeater graph state of $2m$ photons displayed in (c). The boxed area of the circuit is repeated multiple times to generate photons $p_4,p_5,p_6,\cdots, p_{2m-6}$. 
	}
	\label{fig:RGS2m}
\end{figure}

\section*{Supplementary Note 4}

\begin{figure}[t]
	\centering
	\includegraphics[width=0.5\linewidth]{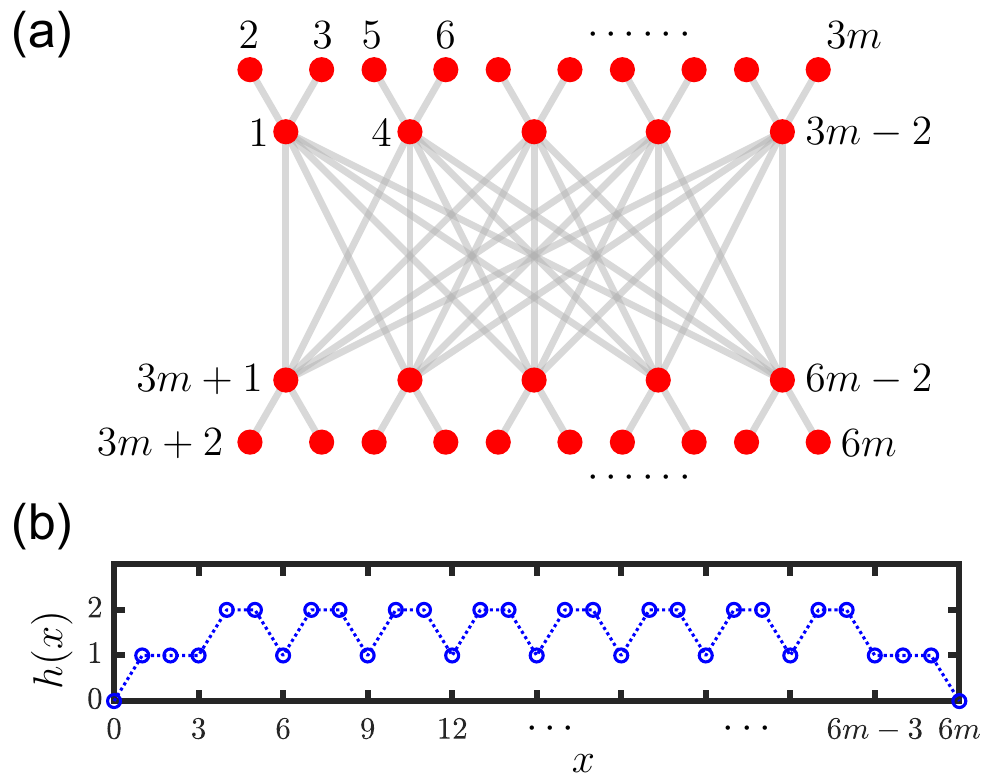}
	\caption{\textbf{Modified RGS example.} (a) The graph for RGS, which has $6m$ vertices. The labels represent the emission sequence. (b) The height function $h(x)$ for the target graph state in (a).
	}
	\label{fig:RGS6m}
\end{figure}

Finally, we consider a modified repeater graph state that includes some additional redundancy to further boost the likelihood of successful Bell measurements \cite{Buterakos_2017PRX_S}.
Supplementary Fig.~\hyperref[fig:RGS6m]{\ref*{fig:RGS6m}}(a) shows an example of such a repeater graph state with $6m$ photons $(m > 3)$. Note that compared to the state in Supplementary Fig.~\hyperref[fig:RGS2m]{\ref*{fig:RGS2m}}(c), this state contains twice as many external photon arms and is missing those internal edges that are not necessary for the functionality of this state as a repeater.
Supplementary Fig.~\hyperref[fig:RGS6m]{\ref*{fig:RGS6m}}(b) shows that the height function is at most $2$ for any $m > 3$, i.e., only two emitter qubits are needed to generate the state in (a). We list all operations in the generation circuit in Supplementary Eq.~\eqref{eq:MPSsolution}. Denoting the $e_1$-th and the $e_2$-th qubits as the emitter qubits, where $e_1 = 6m+1$ and $e_2 = 6m+2$, the circuit is given by:
\begin{equation}
	\begin{aligned}
		W_0 &= \mathrm{CNOT}_{e_1e_2}H_{e_1}H_{e_2} \;\\
		\eta_j &= 
		\left\{
		\begin{aligned}
			&e_2,\quad 6m-5\le j\le 6m - 3\\
			&e_1,\quad \text{otherwise}
		\end{aligned}
		\right.\\
		U_{p,j} &=
		\left\{
		\begin{aligned}
			&\mathbbm{1},\quad j = 1 \text{ (mod 3)}\\
			&H_j,\quad \text{otherwise}\\
		\end{aligned}
		\right.\\
		U_{e,j} &=
		\left\{
		\begin{aligned}
			&\mathrm{CNOT}_{e_2e_1},\quad j = 3\\
			&H_{e_1},\quad j = 3k \text{ with } 2\le k\le2m\\
			&H_{e_2},\quad j = 6m-3\\
			&\mathbbm{1},\quad \text{otherwise}
		\end{aligned}
		\right.
	\end{aligned}
	,\quad
	\begin{aligned}
		\mu_{j} &=\left\{
		\begin{aligned}
			& e_1,\quad j = 3k \text{ with }  2\le k\le 2m,\\
			&\qquad\qquad\qquad\text{and}\;k\ne 2m -1\\
			& e_2,\quad j =  6m - 3\\
			& \text{not assigned},\quad\text{otherwise}
		\end{aligned}
		\right.\\
		W_{j} &=\left\{
		\begin{aligned}
			& \mathrm{CNOT}_{e_1e_2},\quad j = 3k \text{ with } 2\le k\le 2m-2,\\
			&\qquad\qquad\qquad\text{and}\;k\ne m -1\\
			& H_{e_2}\mathrm{CNOT}_{e_1e_2},\quad j = 3m - 3\\
			& H_{e_2},\quad j =  6m - 3\\
			& H_{e_1},\quad j = 6m\\
			& \mathbbm{1}, \quad \text{otherwise}
		\end{aligned}\right.
	\end{aligned}.
\end{equation}
In the above circuit, there are $2m-1$ measurements, and $2m - 1$ $\mathrm{CNOT}$ gates. Plugging these operations into Supplementary Eq.~\eqref{eq:MPSsolution} gives the full sequence.


\nocite{}

\begin{thebibliography}{60}
	
	\bibitem{Raussendorf1WQC}
	Robert Raussendorf and Hans~J. Briegel.
	\newblock A one-way quantum computer.
	\newblock {\em Phys. Rev. Lett.}, 86:5188--5191, (2001).
	
	\bibitem{bartolucci2021fusionbased}
	Sara Bartolucci et~al.
	\newblock Fusion-based quantum computation.
	\newblock {\em Preprint at https://arxiv.org/abs/2101.09310}, (2021).
	
	\bibitem{Schlingemann2001}
	D.~Schlingemann and R.~F. Werner.
	\newblock Quantum error-correcting codes associated with graphs.
	\newblock {\em Phys. Rev. A}, 65:012308, (2001).
	
	\bibitem{ShorCode1995}
	Peter~W. Shor.
	\newblock Scheme for reducing decoherence in quantum computer memory.
	\newblock {\em Phys. Rev. A}, 52:R2493--R2496, (1995).
	
	\bibitem{KITAEV20032}
	A.Yu. Kitaev.
	\newblock Fault-tolerant quantum computation by anyons.
	\newblock {\em Ann. Phys.}, 303(1):2--30, (2003).
	
	\bibitem{nielsen_chuang_2010}
	Michael~A. Nielsen and Isaac~L. Chuang.
	\newblock {\em Quantum Computation and Quantum Information: 10th Anniversary
		Edition}.
	\newblock Cambridge University Press, (2010).
	
	\bibitem{Briegel1998}
	H.-J. Briegel, W.~D\"ur, J.~I. Cirac, and P.~Zoller.
	\newblock Quantum repeaters: The role of imperfect local operations in quantum
	communication.
	\newblock {\em Phys. Rev. Lett.}, 81:5932--5935, (1998).
	
	\bibitem{Dur1999}
	W.~D\"ur, H.-J. Briegel, J.~I. Cirac, and P.~Zoller.
	\newblock Quantum repeaters based on entanglement purification.
	\newblock {\em Phys. Rev. A}, 59:169--181, (1999).
	
	\bibitem{Sangouard2011}
	Nicolas Sangouard, Christoph Simon, Hugues de~Riedmatten, and Nicolas Gisin.
	\newblock Quantum repeaters based on atomic ensembles and linear optics.
	\newblock {\em Rev. Mod. Phys.}, 83:33--80, (2011).
	
	\bibitem{Azuma2015}
	Koji Azuma, Kiyoshi Tamaki, and Hoi-Kwong Lo.
	\newblock All-photonic quantum repeaters.
	\newblock {\em Nat. Commun.}, 6(1):6787, (2015).
	
	\bibitem{Muralidharan2016}
	Sreraman Muralidharan et~al.
	\newblock Optimal architectures for long distance quantum communication.
	\newblock {\em Sci. Rep.}, 6(1):20463, (2016).
	
	\bibitem{Gottesman2012}
	Daniel Gottesman, Thomas Jennewein, and Sarah Croke.
	\newblock Longer-baseline telescopes using quantum repeaters.
	\newblock {\em Phys. Rev. Lett.}, 109:070503, (2012).
	
	\bibitem{Degen2017}
	C.~L. Degen, F.~Reinhard, and P.~Cappellaro.
	\newblock Quantum sensing.
	\newblock {\em Rev. Mod. Phys.}, 89:035002, (2017).
	
	\bibitem{Gisin2007}
	Nicolas Gisin and Rob Thew.
	\newblock Quantum communication.
	\newblock {\em Nat. Photonics}, 1(3):165--171, (2007).
	
	\bibitem{Lugiato_2002}
	L~A Lugiato, A~Gatti, and E~Brambilla.
	\newblock Quantum imaging.
	\newblock {\em J. Opt. B: Quantum Semiclass. Opt.}, 4(3):S176--S183, (2002).
	
	\bibitem{Dowling2008}
	Jonathan~P. Dowling.
	\newblock Quantum optical metrology – the lowdown on high-n00n states.
	\newblock {\em Cont. Phys.}, 49(2):125–143, (2008).
	
	\bibitem{Browne2005}
	Daniel~E. Browne and Terry Rudolph.
	\newblock Resource-efficient linear optical quantum computation.
	\newblock {\em Phys. Rev. Lett.}, 95:010501, (2005).
	
	\bibitem{Gao2010}
	Wei-Bo Gao et~al.
	\newblock Experimental demonstration of a hyper-entangled ten-qubit
	schr{\"o}dinger cat state.
	\newblock {\em Nat. Phys.}, 6(5):331--335, (2010).
	
	\bibitem{Li2020}
	Jin-Peng Li et~al.
	\newblock Multiphoton graph states from a solid-state single-photon source.
	\newblock {\em ACS Photonics}, 7(7):1603--1610, (2020).
	
	\bibitem{Nemoto_PRX_2014}
	Kae Nemoto et~al.
	\newblock Photonic architecture for scalable quantum information processing in
	diamond.
	\newblock {\em Phys. Rev. X}, 4:031022, (2014).
	
	\bibitem{Choi_npjQI_2019}
	Hyeongrak Choi, Mihir Pant, Saikat Guha, and Dirk Englund.
	\newblock Percolation-based architecture for cluster state creation using
	photon-mediated entanglement between atomic memories.
	\newblock {\em npj Quantum Information}, 5(1):104, (2019).
	
	\bibitem{Schon2005PRL}
	C.~Sch\"on, E.~Solano, F.~Verstraete, J.~I. Cirac, and M.~M. Wolf.
	\newblock Sequential generation of entangled multiqubit states.
	\newblock {\em Phys. Rev. Lett.}, 95:110503, (2005).
	
	\bibitem{Schon2007PRA}
	C.~Sch\"on, K.~Hammerer, M.~M. Wolf, J.~I. Cirac, and E.~Solano.
	\newblock Sequential generation of matrix-product states in cavity qed.
	\newblock {\em Phys. Rev. A}, 75:032311, (2007).
	
	\bibitem{Lindner_2009PRL}
	Netanel~H. Lindner and Terry Rudolph.
	\newblock Proposal for pulsed on-demand sources of photonic cluster state
	strings.
	\newblock {\em Phys. Rev. Lett.}, 103:113602, (2009).
	
	\bibitem{Schwartz2016}
	I.~Schwartz et~al.
	\newblock Deterministic generation of a cluster state of entangled photons.
	\newblock {\em Science}, 354(6311):434--437, (2016).
	
	\bibitem{Besse_NatCommun_2020}
	Jean-Claude Besse et~al.
	\newblock Realizing a deterministic source of multipartite-entangled photonic
	qubits.
	\newblock {\em Nat. Commun.}, 11(1):4877, (2020).
	
	\bibitem{Economou_2010PRL}
	Sophia~E. Economou, Netanel Lindner, and Terry Rudolph.
	\newblock Optically generated 2-dimensional photonic cluster state from coupled
	quantum dots.
	\newblock {\em Phys. Rev. Lett.}, 105:093601, (2010).
	
	\bibitem{Gimeno-Segovia2019}
	Mercedes Gimeno-Segovia, Terry Rudolph, and Sophia~E. Economou.
	\newblock Deterministic generation of large-scale entangled photonic cluster
	state from interacting solid state emitters.
	\newblock {\em Phys. Rev. Lett.}, 123:070501, (2019).
	
	\bibitem{Buterakos_2017PRX}
	Donovan Buterakos, Edwin Barnes, and Sophia~E. Economou.
	\newblock Deterministic generation of all-photonic quantum repeaters from
	solid-state emitters.
	\newblock {\em Phys. Rev. X}, 7:041023, (2017).
	
	\bibitem{Russo2018}
	Antonio Russo, Edwin Barnes, and Sophia~E. Economou.
	\newblock Photonic graph state generation from quantum dots and color centers
	for quantum communications.
	\newblock {\em Phys. Rev. B}, 98:085303, (2018).
	
	\bibitem{Hilaire2021}
	Paul Hilaire, Edwin Barnes, and Sophia~E. Economou.
	\newblock Resource requirements for efficient quantum communication using
	all-photonic graph states generated from a few matter qubits.
	\newblock {\em {Quantum}}, 5:397, February 2021.
	
	\bibitem{Zhan2020}
	Yuan Zhan and Shuo Sun.
	\newblock Deterministic generation of loss-tolerant photonic cluster states
	with a single quantum emitter.
	\newblock {\em Phys. Rev. Lett.}, 125:223601, (2020).
	
	\bibitem{Borregaard2020}
	Johannes Borregaard et~al.
	\newblock One-way quantum repeater based on near-deterministic photon-emitter
	interfaces.
	\newblock {\em Phys. Rev. X}, 10:021071, (2020).
	
	\bibitem{michaels2021multidimensional}
	Cathryn~P. Michaels et~al.
	\newblock Multidimensional cluster states using a single spin-photon interface
	coupled strongly to an intrinsic nuclear register.
	\newblock {\em Quantum}, 5:565, (2021).
	
	\bibitem{Pichler2017}
	Hannes Pichler, Soonwon Choi, Peter Zoller, and Mikhail~D. Lukin.
	\newblock Universal photonic quantum computation via time-delayed feedback.
	\newblock {\em PNAS}, 114(43):11362--11367, (2017).
	
	\bibitem{Russo_2019}
	Antonio Russo, Edwin Barnes, and Sophia~E Economou.
	\newblock Generation of arbitrary all-photonic graph states from quantum
	emitters.
	\newblock {\em New J. Phys.}, 21(5):055002, (2019).
	
	\bibitem{NJP_9_204_2007}
	M~Van den Nest, W~Dür, A~Miyake, and H~J Briegel.
	\newblock Fundamentals of universality in one-way quantum computation.
	\newblock {\em New J. Phys.}, 9(6):204 -- 204, (2007).
	
	\bibitem{HMP2006}
	Peter H{{\o}}yer, Mehdi Mhalla, and Simon Perdrix.
	\newblock {Resources Required for Preparing Graph States}.
	\newblock In {\em {17th International Symposium on Algorithms and Computation
			(ISAAC 2006)}}, volume 4288 of {\em Lecture Notes in Computer Science}, pages
	638 -- 649, Kolkata, India, (2006).
	
	\bibitem{PhysRevX.7.031016_Nahum}
	Adam Nahum, Jonathan Ruhman, Sagar Vijay, and Jeongwan Haah.
	\newblock Quantum entanglement growth under random unitary dynamics.
	\newblock {\em Phys. Rev. X}, 7:031016, (2017).
	
	\bibitem{PhysRevB.100.134306}
	Yaodong Li, Xiao Chen, and Matthew P.~A. Fisher.
	\newblock Measurement-driven entanglement transition in hybrid quantum
	circuits.
	\newblock {\em Phys. Rev. B}, 100:134306, (2019).
	
	\bibitem{VandenNest2004_PRA}
	Maarten Van~den Nest, Jeroen Dehaene, and Bart De~Moor.
	\newblock Graphical description of the action of local clifford transformations
	on graph states.
	\newblock {\em Phys. Rev. A}, 69:022316, (2004).
	
	\bibitem{Hein2006}
	M.~Hein et~al.
	\newblock Entanglement in graph states and its applications.
	\newblock {\em Preprint at https://arxiv.org/abs/quant-ph/0602096}, (2006).
	
	\bibitem{Gottesman:1998hu}
	Daniel Gottesman.
	\newblock The {H}eisenberg representation of quantum computers.
	\newblock {\em Preprint at https://arxiv.org/abs/quant-ph/9807006}, (1998).
	
	\bibitem{Cramer_NatCommun_2010}
	Marcus Cramer et~al.
	\newblock Efficient quantum state tomography.
	\newblock {\em Nat. Commun.}, 1(1):149, (2010).
	
	\bibitem{Hein2004PRA}
	M.~Hein, J.~Eisert, and H.~J. Briegel.
	\newblock Multiparty entanglement in graph states.
	\newblock {\em Phys. Rev. A}, 69:062311, (2004).
	
	\bibitem{Orus2014}
	Román Orús.
	\newblock A practical introduction to tensor networks: Matrix product states
	and projected entangled pair states.
	\newblock {\em Ann. Phys.}, 349:117--158, (2014).
	
	\bibitem{Audenaert_2005NJP}
	Koenraad M.~R. Audenaert and Martin~B Plenio.
	\newblock Entanglement on mixed stabilizer states: normal forms and reduction
	procedures.
	\newblock {\em New J. Phys.}, 7:170, (2005).
	
	\bibitem{RandomGraphs}
	E.~N. Gilbert.
	\newblock {Random Graphs}.
	\newblock {\em Ann. Math. Stat.}, 30(4):1141 -- 1144, (1959).
	
	\bibitem{Fattal2004}
	David Fattal, Yoshihisa~Yamamoto Toby S.~Cubitt, Sergey Bravyi, and Isaac~L.
	Chuang.
	\newblock Entanglement in the stabilizer formalism.
	\newblock {\em Preprint at https://arxiv.org/abs/quant-ph/0406168}, 2004.
	
	\bibitem{OUM201715}
	Sang il~Oum.
	\newblock Rank-width: Algorithmic and structural results.
	\newblock {\em Discret. Appl. Math.}, 231:15--24, (2017).
	\newblock Algorithmic Graph Theory on the Adriatic Coast.
	
	\bibitem{Massey1978}
	J.~L. Massey.
	\newblock Foundation and methods of channel encoding.
	\newblock {\em Proc. Int. Conf. on Information Theory and Systems}, 65, (1978).
	\newblock (Berlin, Germany, Sept. 1978).
	
	\bibitem{OUM200579}
	Sang il~Oum.
	\newblock Rank-width and vertex-minors.
	\newblock {\em J. Combin. Theory Ser. B}, 95(1):79--100, (2005).
	
	\bibitem{Adler2017}
	Isolde Adler, Mamadou~Moustapha Kant{\'e}, and O-joung Kwon.
	\newblock Linear rank-width of distance-hereditary graphs {I}. {A}
	polynomial-time algorithm.
	\newblock {\em Algorithmica}, 78(1):342--377, (2017).
	
	\bibitem{Jeong2017}
	Jisu Jeong, Eun~Jung Kim, and Sang-il Oum.
	\newblock The “art of trellis decoding” is fixed-parameter tractable.
	\newblock {\em IEEE Trans. Inform. Theory}, 63(11):7178--7205, (2017).
	
\end{thebibliography}

\begin{thebibliography}{1}
	
	\bibitem{MATLAB_CircuitSolver}
	{Bikun Li}.
	\newblock Photon-emission-circuit-solver: Emission circuit solver for photonics
	stabilizer state, (2021).
	\newblock (https://doi.org/10.5281/zenodo.5652105).
	
	\bibitem{Schon2005PRL_S}
	C.~Sch\"on, E.~Solano, F.~Verstraete, J.~I. Cirac, and M.~M. Wolf.
	\newblock Sequential generation of entangled multiqubit states.
	\newblock {\em Phys. Rev. Lett.}, 95:110503, (2005).
	
	\bibitem{ShorCode1995_S}
	Peter~W. Shor.
	\newblock Scheme for reducing decoherence in quantum computer memory.
	\newblock {\em Phys. Rev. A}, 52:R2493--R2496, (1995).
	
	\bibitem{nielsen_chuang_2010_S}
	Michael~A. Nielsen and Isaac~L. Chuang.
	\newblock {\em Quantum Computation and Quantum Information: 10th Anniversary
		Edition}.
	\newblock Cambridge University Press, (2010).
	
	\bibitem{Buterakos_2017PRX_S}
	Donovan Buterakos, Edwin Barnes, and Sophia~E. Economou.
	\newblock Deterministic generation of all-photonic quantum repeaters from
	solid-state emitters.
	\newblock {\em Phys. Rev. X}, 7:041023, (2017).
	
\end{thebibliography}

\end{document}